\documentclass[%
 reprint,
showpacs,preprintnumbers,
 amsmath,amssymb,
 aps,
]{revtex4-1}
\usepackage[pagebackref=false,colorlinks,linkcolor=blue,filecolor=green,urlcolor=blue ,citecolor=blue]{hyperref}
\usepackage{xcolor}

\usepackage{graphicx}
\usepackage{dcolumn}
\usepackage{bm}
\usepackage{amsmath,mathdots} 
\usepackage{changes}
\usepackage{cancel}
\usepackage[title]{appendix}

\begin{document}
 
\title{Topological properties of subsystem-symmetry-protected edge states in an extended quasi-one-dimensional dimerized lattice}
\author{Milad Jangjan and Mir Vahid Hosseini}
 \email[Corresponding author: ]{mv.hosseini@znu.ac.ir}
\affiliation{Department of Physics, Faculty of Science, University of Zanjan, Zanjan 45371-38791, Iran}

\begin{abstract}
We investigate theoretically  the topological properties of dimerized quasi-one-dimensional (1D) lattice comprising of multi legs $(L)$ as well as multi sublattices $(R)$. The system has main and subsidiary exchange symmetries. In the basis of latter one, the system can be divided into $L$ 1D subsystems each of which corresponds to a generalized $SSH_R$ model having $R$ sublattices and on-site potentials. Chiral symmetry is absent in all subsystems except when the axis of main exchange symmetry coincides on the central chain. We find that the system may host zero- and finite-energy topological edge states. The existence of zero-energy edge state requires a certain relation between the number of legs and sublattices. As such, different topological phases, protected by subsystem symmetry, including zero-energy edge states in the main gap, no zero-energy edge states, and zero-energy edge states in the bulk states are characterized. Despite the classification symmetry of the system belongs to $BDI$ but each subsystem falls in either $AI$ or $BDI$ symmetry class.

\end{abstract}

\maketitle

\section {Introduction} \label{s1}
 
Topological states have engaged in many areas of physics especially in several systems involving electrons \cite{TopoElec1,TopoElec2,TopoElec3,TopoElec4}, cold atoms \cite{TopoCold1,TopoCold2}, photons \cite{TopoPhoto1} in past decade \cite{TopoHall}. One of the properties of these states is their robustness against disorders or defects \cite{TopoMat} as long as the fundamental symmetry of system protecting nontrivial topology is preserved. Symmetries of a system can be used to classify the topology of electronic band structures. Based on topological classification \cite{class1,class2,class3,class5,Gclass1,Gclass2,Gclass3}, a diversity of new and interesting materials can be categorized into topological insulating \cite{TopoElec1,TopoIns1, TopoIns2,TopoIns3,TopoIns4}, topological superconducting \cite{TopoElec2,ChernIn,TopoSup1,TopoSup2,TopoSup3}, topological semi-metal \cite{TopoSem1,TopoSem2,TopoSem3,TopoSem4}, and topological metal \cite{TopoMet1,TopoMet2,TopoMet3,TopoMet4, TopoMet5,TopoMet6,TopoMet7} states in one, two, or three dimensions.

In one-dimensional (1D) systems, one of the simplest models for topological insulators is the Su-Schrieffer-Heeger (SSH) chain \cite{SSH}. It is characterized by two different tunneling amplitudes, i.e., intra- and inter-unit cell tunneling amplitudes, between two different sublattices in a chain \cite{SSH1}. The SSH chain can host localized edge states at its ends under open boundary conditions in the topologically nontrivial regime. Such topological state is protected by inversion symmetry and has its topological invariant that can be calculated via bulk states of the band structure under periodic boundary conditions \cite{SSH2}. The generalization of SSH model including next nearest hopping \cite{GeneralizedSSH}, spin-orbit interaction \cite{SSHSpinOrbit,SpinZem} and Zeeman field \cite{SpinZem} has also been studied recently. Furthermore, the extended SSH chain, with more than two sublattices per unit cell, comprising three \cite{SSH3} and four \cite{SSH40,SSH4} sublattices, or even \cite{GSpinZem} and odd \cite{TopoMet6} numbers of sublattices per unit cell has been investigated in 1D geometry revealing various topological phases protected by main symmetries of the whole system. 

On the other hand, there are interesting 2D systems hosting topological phases \cite{2DTI,2DTI1}. Chern insulator has been proposed in a 2D lattice exhibiting a nonzero quantized Hall conductance in the absence of an external magnetic field \cite{ChernIn,TopoSup1}. Furthermore, helical edge states, protected by time-reversal symmetry, have been realized at the edge of 2D HgTe quantum wells \cite{TopoSup2}. Interestingly, it has been shown that nontrivial topological phases can be hosted in a 2D SSH model in the absence of Berry curvature due to presence of both time-reversal and inversion symmetries \cite{Topo2D}.  

However, in going from one dimension to two dimensions, one deals with a class of systems, namely, quasi-1D systems, having spectacular features \cite{Quasi1DReview}. Simple examples of quasi-1D systems are coupled chains and ladders. Recently, several topological features of ladder lattice structures have been investigated exhibiting a rich variety of phases based on their topological properties, for instance, topological superconductivity in Kitaev ladder \cite{KitaevLadder}. Also, the effects of spin-orbit coupling \cite{SpinOrbitCreutzLadder}, interactions \cite{InteractionsCreutzladder}, and interchain coupling \cite{interchainCoupCreutzLadder} on the topological feature of Creutz ladder as well as the role of topology on the charge pumping phenomenon in the Creutz ladder \cite{chargePumpCreutzLadder} have been studied. Beyond the single SSH chain, in a two-leg SSH chain, topological nodal points \cite{TopoNodalPoints} and non-Abelian Berry connections associated with the glide reflection symmetry \cite{GlideRefl} have been investigated. Also, chiral solitons \cite{ChiralSolitonExp} and topological bound states \cite{TopoBoundExp} have been observed in a coupled double SSH chain. It has also been shown that a dimerized two-leg ladder can host localized and delocalized topological finite-energy edge states in continuum for asymmetric and symmetric dimerization patterns, respectively \cite{TopoMet4}. However, topological characterization of generalized ladder systems involving more than two legs with multi sublattices per leg deserves to be explored further based on quasi-1D materials $Bi_4X_4$ ($X=Br,I$) \cite{Quasi1DTI,Quasi1DTI2}. Also, it is interesting to find a situation where topological phases are not protected by main symmetries of the whole system.  

In this work, we consider a quasi-1D dimerized lattice consisting of $L$ legs such that each leg has $R$ sublattices. We will investigate topological properties of the system according to its symmetries. The simultaneous existence of reflection and (main and subsidiary) inversion symmetries allows to define the so-called main and subsidiary exchange symmetries. The Hamiltonian of the system can be block-diagonalized in the basis of subsidiary exchange symmetry with $L$ blocks. Each block can be regarded as a 1D subsystem that resembles the SSH model with $R$ sublattices, $SSH_R$, having an effective on-site potential. In the topological regime, there are zero- and finite-energy edge states. The system hosts $m-1$ zero-energy edge states out of $L(R-1)$ edge states where $m$ is the greatest common divisor of ($L+1,R$). Subsequently, depending on the values of $L$ and $R$, we realize different topological phases: i) zero-energy edge states reside in the main gap, ii) there are no zero-energy edge states, and iii) zero-energy edge states reside within bulk states. By breaking symmetries of the system, interestingly, we realize that the inversion symmetry of subsystems protects the topology of the system.

\section {Model and Theory} \label{s2}

\begin{figure}[t!]
\centerline{\includegraphics[width=8.5cm]{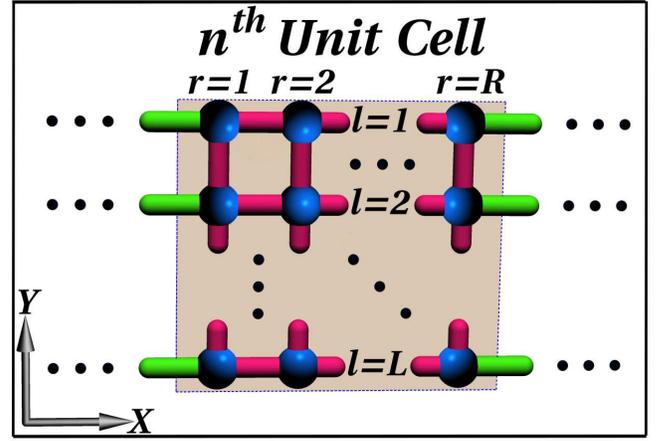}}
\caption{(Color online) Schematic geometry of a finite section of quasi-1D lattice comprised of $L$ coupled chains oriented along the $x$ direction. Each chain has $R$ sublattices. Intra- (inter-) unit cell hopping, $t$ ( $t^\prime$), represented in red (green) color.}
\label{fig1}
\end{figure}

We start by considering a quasi-1D superlattice comprising of $L$ number of chains along the x-direction and each chain contains $R$ number of sublattices as shown in Fig. \ref{fig1}. The Hamiltonian of the system being the sum of the Hamiltonian of chains, $H_{chain}$, and the Hamiltonian of inter chain couplings, $H_{coupling}$, is 
\begin{eqnarray}\label{RealHomil}
 H&=&H_{chain} + H_{coupling},  \\
 H_{chain}&=&\sum_{n=1}^N\sum_{l=1}^{L}\sum_{r=1}^{R-1} t_l C_{n,l,r}^\dagger C_{n,l,r+1} \nonumber \\
 & +& \sum_{n=1}^{N-1}\sum_{l=1}^{L} t_l^\prime C_{n,l,R}^\dagger C_{n+1,l,1}+h.c, \nonumber \\
 H_{coupling}&=&\sum_{n=1}^N\sum_{l=1}^{L-1}\sum_{r=1}^R t_r C_{n,l,r}^\dagger C_{n,l+1,r}+h.c,\nonumber
\end{eqnarray}
where $C^{\dagger}_{n,l,r}$ ($C_{n,l,r}$) is the creation (annihilation) operator of electron on the $r$th sublattice of the $l$th chain at the $n$th unit cell. We take the intra- (inter-) cell hopping $t_l=t=1+\delta_0\cos(\theta)$ ($t_l^\prime=t^\prime=1-\delta_0cos(\theta)$) $\forall l \in [1, L]$ in each leg and inter leg hopping $t_r=t=1+\delta_0\cos(\theta)$ $\forall r \in [1, R]$ with $\delta_0$ and $\theta$ being the dimerization amplitude and a cyclically varying parameter to control the strength and sign of dimerization, respectively. Also, the $N$ is the number of unit cells. Without loss of generality, we take $\delta_0=0.8$ throughout the paper.

\begin{figure*}[t!]
    \centering
    \includegraphics[width=1\linewidth]{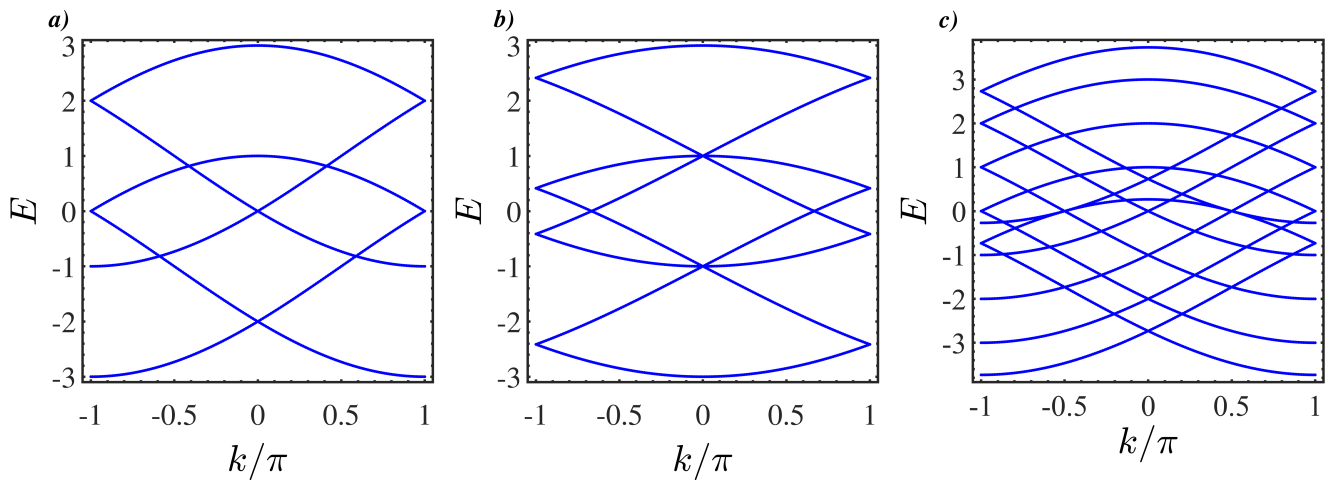}
    \caption{(Color online) Energy spectra under periodic boundary conditions for (a) $(L,R)=(2,3)$, (b) $(L,R)= (2,4)$, and (c) $(L,R)=(5,3)$. Here $\theta/\pi=1/2$.}
    \label{fig2}
\end{figure*}

Assuming periodic boundary conditions along the chains and using Fourier transform of annihilation (creation) operator $C^{(\dagger)}_{n,l,r}=1/\sqrt(N)\sum_k e^{(-)ikn}C^{(\dagger)}_{k,l,r}$, we can write Hamiltonian (\ref{RealHomil}) in reciprocal space as $H=\sum_k \langle\psi_k| h(k)|\psi_k\rangle$ where $|\psi_k\rangle=\sum_{l,r}^{L,R} C_{k,l,r}|l \rangle \otimes |r\rangle$ and
\begin{eqnarray}\label{reciprocalHomil}
&h&(k)=\begin{pmatrix}
 h_{chain}& h_{coupling}& \ldots &0\\
 h_{coupling} & h_{chain} &\ddots& 0&\\
  \vdots & \ddots &\ddots&h_{coupling}\\
  0 & \ldots & h_{coupling} & h_{chain}
   \end{pmatrix}_{L\times L}, 
\end{eqnarray}
with
\begin{eqnarray}
&h&_{chain}=\begin{pmatrix}
  0 &  t  & \ldots & t^\prime e^{ik}\\
  t & &  \ddots& \vdots\\
   \vdots & \ddots & &t \\
   t^\prime e^{-ik} & \ldots &t&0\\
 \end{pmatrix}_{R\times R}, \nonumber\\
&h&_{coupling}=\begin{pmatrix}
 t &  0  & \ldots & 0\\
 0 & t&  \ddots& \vdots\\
 \vdots & \ddots &t &0 \\
 0 & \ldots &0&t\\
\end{pmatrix}_{R\times R}.\nonumber
\end{eqnarray}

Diagonalizing the bulk Hamiltonian (\ref{reciprocalHomil}), the spectrum can be obtained having $LR$ bands as shown in Fig. \ref{fig2} for different values of $L$ and $R$ with $\theta/\pi = 1/2$. One can see that there are some band touching points at the symmetric points $k = 0$ and $k =\pi$ for $\theta/\pi = 1/2$ (and for $\theta/\pi =3/2$ not shown). Interestingly, the gap closings at the symmetric points, which may be the signal of occurring topological phase transition, take place between certain bands at zero and/or finite energies depending on the values of $L$ and $R$. This indicates that there would be possibly zero- and/or finite-energy edge states under open boundary conditions. As shown in Fig. \ref{fig2}(a), the topological band touchings can occur between different bands at zero energy while bulk states do not cross the Fermi level, $E=0$, predicting that the zero-energy edge states will be resided in the main gap. In contrast, as shown in Fig. \ref{fig2}(b), the topological band touchings only exist at finite energies and the bulk states of different bands can be accessed at zero energy. Consequently, in this case, there will be no zero-energy edge states. From Fig. \ref{fig2}(c), one can see, in addition of topological band touching points at zero energy, bulk states of the other bands are available with the same energy. So, one may anticipate that the zero-energy edge states and the bulk states will coexist. Therefore, these features imply that, correspondingly, there would be possibly symmetry-protected topological phases associated with zero-energy edge states in the gap, without zero-energy edge states, and with zero-energy edge states in the bulk states. So, we are interested in inspecting symmetries of the system in the following.

\section{Symmetries of the system}
\label{s3}


\subsection{Main symmetries}


The system exhibits time-reversal symmetry defined as $\mathcal{T}_{i} h(k)\mathcal{T}_{i}=h(-k)$ with $i=1,2$. The operators of time-reversal symmetry are $\mathcal{T}_1= \tau_{x_R} \otimes I_L \mathcal{K}$ and $\mathcal{T}_2=I_{LR} \mathcal{K}$ with $\mathcal{K}$ being the complex conjugation, $I_L$ ($I_{LR}$) is an identity matrix of size $L$ ($LR$) and

\begin{eqnarray}\label{e3}
 \tau_{x_R}&=&\begin{pmatrix}
 & &  & &1\\
  & O& &1 &\\
  & &\iddots & & \\
   & 1&  &O & \\
   1 & & & & \\
 \end{pmatrix}_{R\times R}.
\end{eqnarray}

Also, the system shows particle-hole and chiral symmetries. Since the system under consideration is a spineless system, the particle-hole operator, $\mathcal{P}$, and the chiral operator, $\Gamma$, are identical: $\mathcal{P}_1=\Gamma_1=\mathcal{C}_{1_L}\otimes\mathcal{C}_{1_R}$ and $\mathcal{P}_2=\Gamma_2=\mathcal{C}_{2_L}\otimes\mathcal{C}_{1_R}$ where 
\begin{eqnarray}\label{e6}
 \mathcal{C}_{1_D}=\begin{pmatrix}
 1& &  & &\\
  & -1& &O &\\
  & &\ddots & & \\
   & O&  & & \\
    & & & & \\
 \end{pmatrix}_{D\times D},
 \end{eqnarray}
and
\begin{eqnarray}\label{e3}
 \mathcal{C}_{2_D}&=&\begin{pmatrix}
 & &  & &1\\
  & O& &-1 &\\
  & &\iddots & & \\
   & &  &O & \\
    & & & & \\
 \end{pmatrix}_{D\times D}.
\end{eqnarray}

The number of sublattices in each chain can be either odd or even, so there exist two types of formula for each symmetry. For the even number of sublattices per chain, the particle-hole and chiral symmetries satisfy the general relations $\mathcal{P}_ih(k)\mathcal{P}_i=-h^{\star}(-k)$ and $\Gamma_i h(k)\Gamma_i=-h(k)$ with $i=1,2$, respectively. When the $R$ gets odd numbers, there are hidden particle-hole and chiral symmetries fulfilling $\mathcal{P}_ih(k)\mathcal{P}_i=-h^{\star}(-k-\pi)$ and $\Gamma_i h(k)\Gamma_i=-h(k-\pi)$, respectively. Note that all the operators of each symmetry can be commuted with each other as $[\boldsymbol{\Sigma}_i,\boldsymbol{\Sigma}_j]=0$ where $\boldsymbol{\Sigma}$=($\mathcal{T}, \mathcal{P}, \Gamma$).

Due to $\mathcal{T}^2_i=\mathcal{P}^2_i=\Gamma^2_i=I_{LR}$, according to the primary topological periodic table \cite{class1,class2,class3,class5}, which is based on nonspatial symmetries, the topological class of the system falls into $BDI$ class with $\mathbb{Z}$ index and the band structure of the system may be gapped near the Fermi level revealing degenerate edge states in topologically nontrivial phases.

Furthermore, Hamiltonian (\ref{reciprocalHomil}) illustrates reflection and inversion symmetry. Although in quasi-1D systems, both inversion and reflection symmetries change $k \rightarrow-k$, they could have a different form of operators. The reflection and inversion symmetry defining $\Pi_i h(k) \Pi_i=h(-k)$ have the operators $\Pi_1=I_R \otimes \tau_{x_R}$ and $\Pi_2=\tau_{x_L}\otimes \tau_{x_R}$, respectively. Note that the inversion symmetry can be regarded as the reflection symmetry for each chain. The mirror line of reflection symmetry is perpendicular to the orientation of chains and bisects the system while the inversion symmetry has an inversion point located in the center of the system. 

In addition to the main inversion symmetry, already discussed above, there is a subsidiary inversion symmetry whose inversion points can be placed between every two adjacent chains at the mirror line. The operator of subsidiary inversion symmetry can be found as   
\begin{eqnarray}\label{e3}
 \Pi_3=\mathcal{C}_{3_L} \otimes \tau_{x_R},
\end{eqnarray}
where 
 \begin{eqnarray}\label{e6}
 \mathcal{C}_{3_L}=\begin{pmatrix}
  0 &  1  &  &O\\
  1 &\ddots &  \ddots& \\
    & \ddots & \ddots &1 \\
   O &  &1&0\\
 \end{pmatrix}_{L\times L}.
 \end{eqnarray}
It should be noted that the subsidiary inversion symmetry exists provided that the main inversion symmetry is preserved.   
\subsection{Additional symmetries}

In this system, as already mentioned, both the inversion and reflection symmetries can change $k$ to $-k$, therefore, under a transformation that is a combination of the inversion with respect to the main inversion point and reflection with respect to the vertical mirror line, the Hamiltonian is invariant. This implies that there is an additional symmetry, namely, exchange symmetry, with a horizontal mirror line, along the $x$ axis, in the middle of the lattice. Correspondingly, its exchange operator can be obtained by multiplying the reflection operator $\Pi_1$ by the inversion operator $\Pi_2$, i.e., 
\begin{eqnarray}\label{e5}
 \Upsilon_1 =\Pi_1\Pi_2=\tau_{x_L} \otimes I_R,
\end{eqnarray}
exchanging a chain from the upper half with its corresponding one in the lower half of the system. Also, the existence of the subsidiary inversion symmetry enables us to define a subsidiary exchange symmetry. Its operator is the matrix multiplication of $\Pi_2$ and $\Pi_3$, 
\begin{eqnarray}\label{e4}
 \Upsilon_2 =\Pi_3\Pi_2=\mathcal{C}_{3_L}\otimes I_R,
\end{eqnarray}
exchanging two adjacent chains with respect to an axis, being parallel to the chains, located in the middle between them. Note, there would be $ \Upsilon_2$ as long as $ \Upsilon_1$ is established. 

After obtaining the symmetries that leave the Hamiltonian invariant, it is possible to utilize them in dividing the system into subsystems by block-diagonalizing the system Hamiltonian \cite{OddCh1,OddCh2}. This makes it easy to inspect  topological origins of system ingredients, in particular, when a system hosts finite-energy topological edge states \cite{TopoMet7}. Therefore, because of $[h(k), \Upsilon_{i}]=0$ ($i=1,2$), Hamiltonian (\ref{reciprocalHomil}) can be brought into a block-diagonal form through a unitary transformation $\mathcal{H}(k)=Uh(k)U^{-1}$. The unitary matrix $U$ can be constructed from the eigenstates of $\Upsilon_{i}$. According to Eq. (\ref{e5}), it is easy to show that the eigenvalues of $\Upsilon_{1}$ are $\pm1$. So, in the basis of $\Upsilon_{1}$, the Hamiltonian will be block diagonalized into two decoupled blocks. If $L$ is an even number, the size of both blocks is $LR/2$. While for odd $L$, the sizes of two blocks are $\left \lfloor \frac{L}{2} \right \rfloor R$ and $(1+\left \lfloor \frac{L}{2} \right \rfloor)R$. In this case, one should find another exchange symmetry for each block and repeat the block-diagonalization process to obtain $L$ blocks of size $R$. 

On the other hand, the subsidiary exchange symmetry operator $\Upsilon_2$ has $L$ eigenvalues, $\lambda_l$, with $R$-fold degeneracy. The simple closed-form for the eigenvalues $\lambda_l$ is \cite{TridiMatrix1,TridiMatrix2}
\begin{eqnarray}\label{e4}
 \lambda_l = 2 cos(\frac{l\pi}{L+1}), \quad l=1,2,...,L.
\end{eqnarray}
So, in the $\Upsilon_2$ representation, the Hilbert space of the system can be decomposed into $L$ subspaces giving rise the Hamiltonian as $\mathcal{H}(k)=\bigoplus_{l=1}^L h_{\Upsilon_{2}=\lambda_l}(k)$ where the Hamiltonian of each subsystem is
\begin{eqnarray}\label{e7}
h_{\Upsilon_{2}=\lambda_{l}}(k)&=&\begin{pmatrix}
  t\lambda_{l} &  t  & \ldots & t^\prime e^{-ik}\\
  t &t\lambda_{l} &  \ddots& \vdots\\
   \vdots & \ddots & \ddots &t \\
   t^\prime e^{ik} & \ldots &t&t\lambda_{l}\\
 \end{pmatrix}_{R \times R}. 
\end{eqnarray}
Note that the Hamiltonian of subsystem (\ref{e7}) is a Hamiltonian of the extended $SSH$ model \cite{TopoMet6,SSH3,SSH4,GSpinZem}, i.e., $SSH_R$, including $R$ sublattices, with diagonal entries. Evidently, the multiplication of the coupling term $t$ and the subsidiary exchange operator eigenvalues $\lambda_l$ creates \emph{effective on-site potentials} (see Eq. (\ref{e7})). In analog to the original $SSH$ chain model with $R=2$, where there is only one gap, being the main gap, containing one pair of topological edge states in the topological regime, the subsystem $SSH_R$ has $R-1$ gaps, including a main gap and subgaps. Each gap would host topological edge states. Consequently, in the present system with $L$ subsystems, there could be a total of $L(R-1)$ topological edge states with zero and/or finite energies.

It is worthwhile noting that if the $L$ takes an even number, the system Hamiltonian is similar to an $L/2$-spinfull 1D system with $SSH_R$ chain that is exposed to an external Zeeman field with amplitude $t\lambda_l$. In this case, the exchange symmetry is equivalent to the spin-rotation symmetry in a 1D spinfull system \cite{TopoMet2}. For odd $L$, the system resembles an $L$-chain bosonic system with an integer spin $S=\lfloor L/2 \rfloor-1$ \cite{BosonLadder}.

As mentioned above, the topological phase transition occurs at $k=0, \pi$ and $\theta/\pi=1/2, 3/2$. By substituting these requirements in Eq. (\ref{e7}), the energy bands at the phase transition points can be obtained as \cite{TridiMatrix2}
\begin{eqnarray}\label{e12}
    E_{lr} &=& t [\lambda_l + 2 cos(\frac{2 r\pi-\pi}{R})], \quad r=1,2,...,R, \ \textrm{if}\ k=\pi, \nonumber \\
  E_{lr} &=& t [\lambda_l + 2 cos(\frac{2 r\pi}{R})], \quad r=1,2,...,R, \ \textrm{if}\ k=0.
\end{eqnarray}
Subsequently, using Eqs. (\ref{e4}) and (\ref{e12}), gap closure conditions, i.e., $E_{lr}=0$, implies that 
\begin{eqnarray}\label{e13}
k&=&0 \rightarrow 2r = \frac{\pm R l}{1+L}+R, \label{e13}\\
k&=&\pi \rightarrow 2r-1= \frac{\pm R l}{1+L}+R. \label{e130}
\end{eqnarray}
Note, for $k=0$ ($\pi$) the left hand side of the above relation is a positive even (odd) quantity, since the $r$ takes positive integer values. Also, for given integers $R$ and $L$, the values of $l$ satisfying in Eqs. (\ref{e13}) and  (\ref{e130}) gives us the subsystems that their gap can be closed and reopened at zero energy, namely, the main gap of the whole band structure. It is easy to show (see Appendix \ref{AppxA}) that for such subsystems with the corresponding $\lambda_l$, the index $l$ fulfills the following relation 
\begin{eqnarray}\label{e15}
\frac{L+1}{m} \le \ l \le \ (m-1)\frac{L+1}{m},
\end{eqnarray}
where $m$ is the greatest common divisor of ($L+1,R$). From Eq. (\ref{e15}), one finds that there are $m-1$ subsystems for which main gap closing/reopening can take place. After topological phase transition, the gap of each of these subsystems can contain one pair of zero-energy edge states. As a consequence, in total, there will be $m-1$ topological edge states at zero energy under open boundary conditions. 

\section{Symmetries of the subsystems} \label{s4}

\subsection {Odd numbers of L } 

In this case, the axis symmetry of the main exchange symmetry, $\Upsilon_1$, coincides with the central chain and, at the same time, the number of eigenvalues of $\Upsilon_2$, i.e., $\lambda_l$, will be an odd number. As a result, one of the eigenvalues $\lambda_l$ must be zero originating from the fact that eigenvalues of a Hermitian operator, e.g., $\Upsilon_2$, are symmetric about zero. This means that one of the subsystems whose effective on-site potential is zero, $t\lambda_l=0$, has chiral symmetry. Such subsystem reminisces of the bare $SSH_R$ system. However, the whole Hamiltonian has chiral symmetry. Depending on $R$ that would take even or odd numbers, the system hosts topological edge states within either gapped or gapless bulk states around zero energy.

In fact, the even number of sublattices provides bulk insulating ground states supporting particle-hole and chiral symmetries in 1D systems \cite{SSH,SSH1,SSH2,SSH40,SSH4,GSpinZem}. Also, a pair of edge states manifests itself at zero energy because of the presence of inversion and chiral symmetry of the bare $SSH_R$ subsystem \cite{SSH4,GSpinZem}. So, the system at least hosts a pair of zero-energy edge states in a topological insulating regime. In the present cases, the chiral operators commute with the subsidiary exchange operator $\Upsilon_2$ when the $L$ takes odd numbers, i.e., $[\Gamma_i,\Upsilon_2]=0$. As a consequence, according to Refs. \cite{class1,class2,class3}, the class of the subsystem with $\lambda_l=0$ (the subsystem with zero energy edge states) remains $BDI$ hosting nontrivial topological phases.

While the other subsystems lack chiral symmetry due to $\lambda_l \ne 0$. But, as already discussed above, these subsystems would have zero-energy edge states if Eqs. (\ref{e13})-(\ref{e15}) established. In such a situation, $m >1$ and the effective on-site potential forces the energy of finite-energy edge states of the corresponding subsystems to be shifted towards zero energy. This is in contrast to the usual cases where there is no zero-energy topological edge state when there is no chiral symmetry \cite{TopoMet6}. Also, the topological classification of the subsystems without chiral symmetry belongs to $AI$ class owing to the presence of inversion and time-reversal symmetries in the subsystems \cite{Gclass1,Gclass2,Gclass3}.

On the other hand, an odd number of $R$ providing an odd number of bands in each subsystem results in bulk metallic ground states, due to the existence of chiral symmetry and the subsystem with $\lambda_l=0$ resides always within the gapless bulk states. Moreover, if the conditions (\ref{e13})-(\ref{e15}) are established in the other subsystems $\lambda_l \ne 0$, their zero-energy topological edge states would lie in the continuum instead of band gaps.

\subsection {Even numbers of L} 
For an even number of chains, none of the eigenvalues of subsidiary exchange symmetry is zero. So all the subsystems lack chiral symmetry though each chain may include either even or odd number sublattices. This implies that, unlike the previous case, there is no even one subsystem belonging to $BDI$ class. For the even or odd number of $R$, all the subsystems have $AI$ class because of inversion and time-reversal symmetries in the subsystems \cite{Gclass1,Gclass2,Gclass3}. Moreover, the subsystems can be considered as the generalized $SSH_R$ and have finite-energy edge states unless Eqs. (\ref{e13})-(\ref{e15}) would be held for some of them. 


\begin{figure}[t]
    \centering
    \includegraphics[width=1\linewidth]{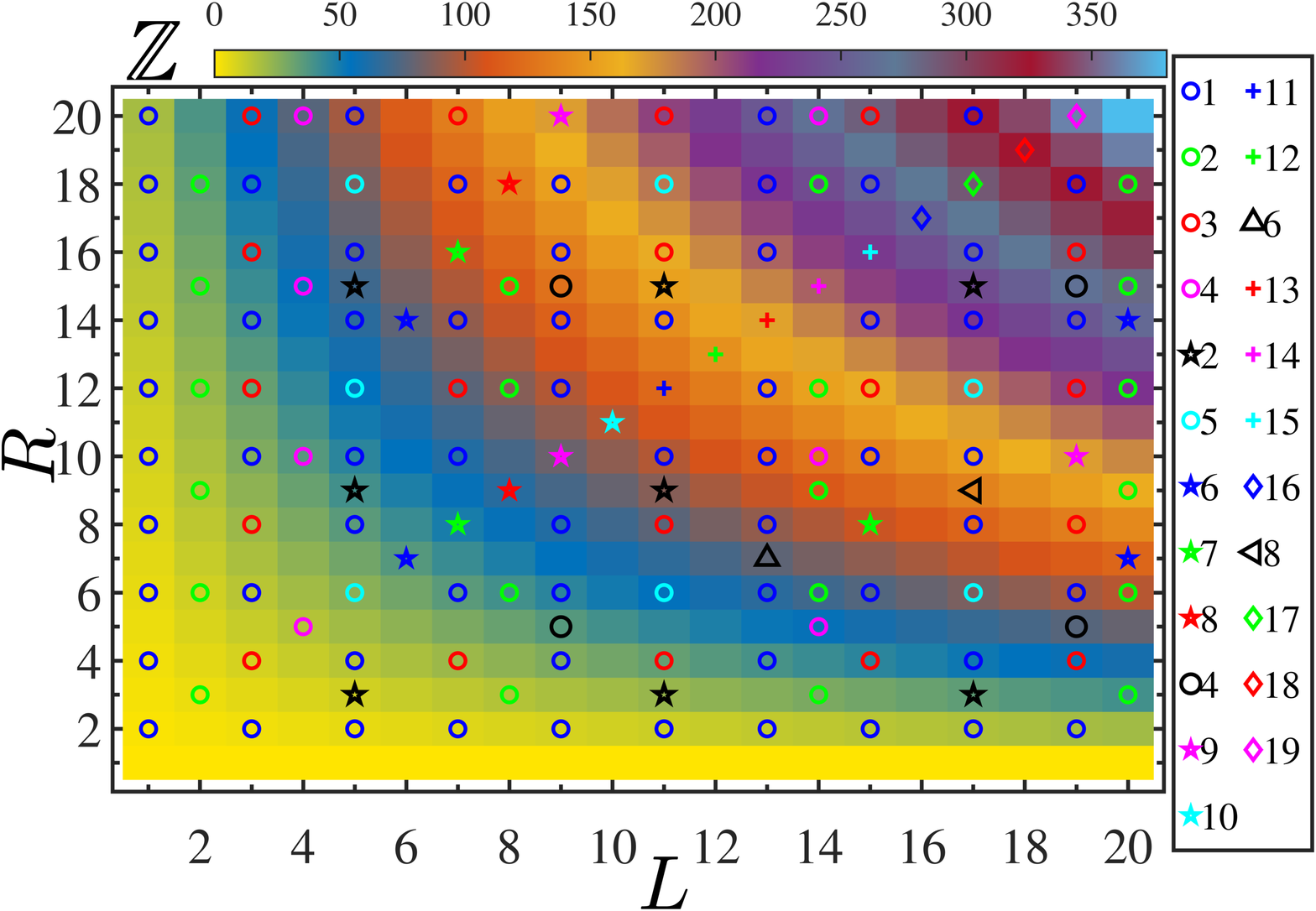}
    \caption{(Color online) Topological phase diagram of the system as functions of the chain number $L$ and the sublattice number $R$. The color indicates the total numbers of edge states ($\mathbb{Z}$). Also, the colored and the black markers represent the number of edge state in the main gap and in the continuum at zero energy, respectively.}
    \label{fig3}
\end{figure}

\begin{figure*}[t]
    \centering
    \includegraphics[width=1\linewidth]{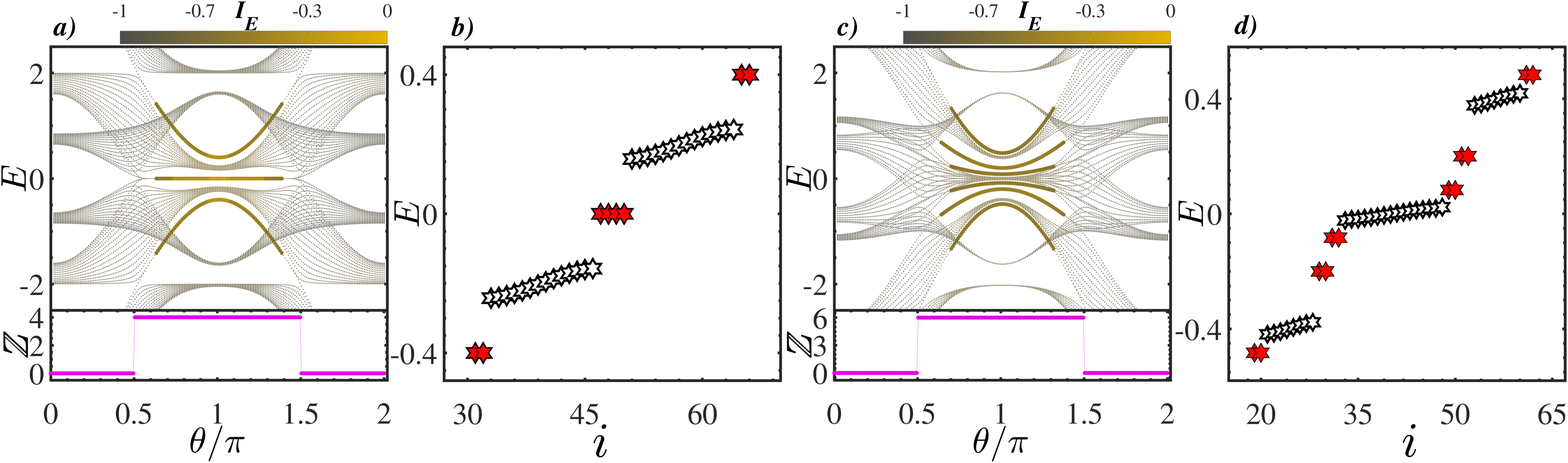}
    \caption{(Color online) Energy spectra and relevant topological invariant $\mathbb{Z}$ as a function of $\theta/\pi$ under open boundary conditions for $L=2$ with (a) $R=3$ and (c) $R=4$. The edge states appear in yellow and the extended bulk ones in gray color. (b) and (d) Energy spectrum as a function of wave function index corresponding to panels (a) and (c), respectively, with $\theta/\pi=1$. Edge and bulk states are shown by red and white stars, respectively.}
    \label{fig4}
\end{figure*}

\subsection {Topological invariant}

The existence of inversion symmetry in the subsystems gives rise that their Hamiltonians commute with the inversion symmetry operator at $k=0$ and $k=\pi$. As such, the subsystems have well-defined parities. Also, the expectation value of inversion operator is $\pm 1$ at either $k=0$ or $k=\pi$. We can define topological invariant $\mathbb{Z}=\sum_{l,r}^{L,R-1}|n_{0,l,r}-n_{{\pi},l,r}|$ \cite{ZInvariant} where $n_{0,l,r}$ and $n_{{\pi},l,r}$ are the number of negative parities at $k=0$ and $k=\pi$, respectively, for the $lth$ subsystem and $rth$ gap. 

Actually, in order to distinguish localized and extended features of an eigenstate $\psi_E$ in the corresponding eigenenergy $E$ under open boundary conditions, we determine inverse participation ratio \cite{IPR} as 
\begin{align}\label{e16}
I_E=\frac{Ln \sum_j \vert \psi_E(j) \vert ^4}{Ln LRN}.
\end{align}
In the above relation, the system would host localized and extended states, respectively, for $I_E=0$ and $I_E=-1$. 

\section {Phase diagram and band structures} \label{s5}

In Fig. (\ref{fig3}), we have plotted the phase diagram of system as function of $R$ and $L$. The total numbers of topological edge states, $\mathbb{Z}$, at both zero and finite energies are represented by colors. According to Eq. (\ref{e15}), we specified the number of zero-energy edge states in the main gap by the different colorful markers. Since for an odd number of $R$ and $L$ the system may host topological edge states in the continuum, we distinguish the number of edge states at zero energy in the continuum by black markers. One can see that the $\mathbb{Z}$ increases rapidly when both $R$ and $L$ increase rather that one of the parameters gets fixed values. Also, for a given $L$ there are at most $L$ zero-energy edge states in the main gap, if $R=p(L+1)$ where $p=1,2,3,...$ . 

The band structure and the related topological invariant $\mathbb{Z}$ as a function of $\theta$ with $L=2$ are shown in Figs. \ref{fig4}(a) and \ref{fig4}(c), respectively, for $R=3$ and $R=4$. In these cases, our system reduces to a two-leg ladder with three and four sublattices per leg. Because, here, the main exchange symmetry axis does not coincide on any chain, there is no zero eigenvalue for the subsidiary exchange symmetry operator $\Upsilon_2$. Also, the whole system can be considered as a direct sum of two decoupled subsystems (or chains) with effective on-site potentials. 

One can see in Fig. \ref{fig4}(a) that there are $L(R-1)=4$ topological edge states in the topological regime with $\mathbb{Z}=4$. Also, according to $(L+1) = 3$ and $R=3$, the great common divisor is $m=3$, and therefore there exist $m-1=2$ pairs of zero-energy edge states in the main gap which are protected by inversion symmetry of their subsystem Hamiltonians. In this case, the effective on-site potential, originated from the combination of coupling term and eigenvalue of $\Upsilon_2$, leads to shifting the finite-energy edge states to zero energy. Furthermore, the remaining $L(R-1)-(m-1)=2$ topological edge states are finite-energy edge states within subgaps. Interestingly, for certain values of $\theta$ the finite-energy edges states cross the bulk states. In Fig. \ref{fig4}(b), the energy spectrum of the system versus wave function index corresponding to the panel (a) is presented for $\theta/\pi=1$. One can see that there are two (two) pairs of red stars at zero (finite) energy showing the number of nontrivial midgap zero- (finite-) energy edge states. 

For the parameters of Fig. \ref{fig4}(c), we have $(L+1)=3$ and $R=4$ thus $m=1$. As a result, the change of sublattice numbers causes the main gap to be closed. Subsequently, there is no zero-energy edge state, $m-1=0$, and the total number of emerged edge states, $L(R-1)=6$, would be hosted as finite-energy topological ones with $\mathbb{Z}=6$ in the topological regime. Correspondingly, as can be seen from Fig. \ref{fig4}(d), there are only six pairs of edge states, represented by red stars, distributed at different finite energies.

 \begin{figure}[t!]
    \centering
    \includegraphics[width=1\linewidth]{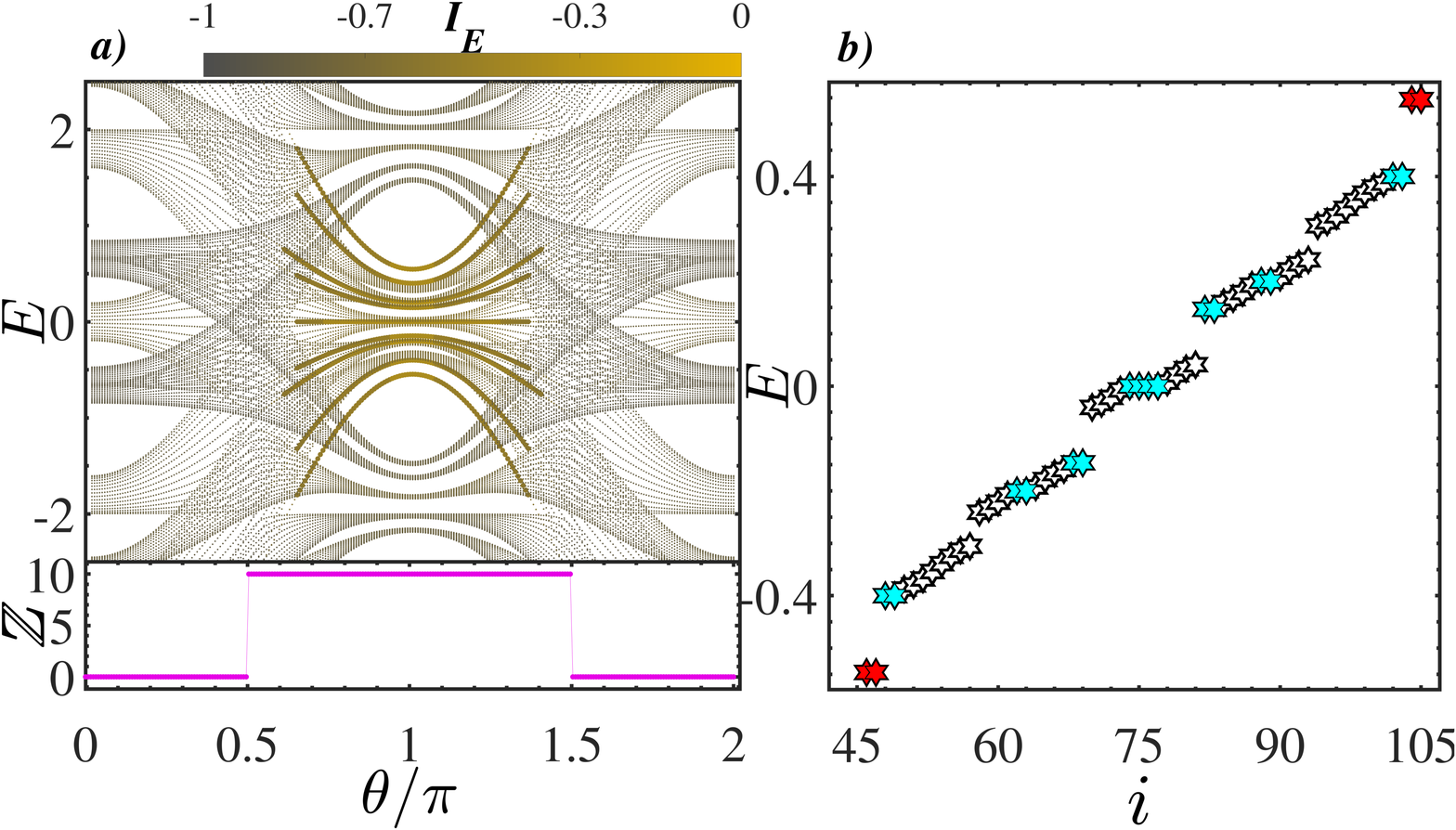}
    \caption{(Color online) (a) Energy spectra and relevant topological invariant as a function of $\theta/\pi$ under open boundary conditions with $(L,R)=(5,3)$. The edge states are shown in yellow and the extended ones in gray color. (b) Energy spectrum as a function of wave function index corresponding to panel (a) with $\theta/\pi=1$. The red, blue, and white markers represent, respectively, the edge states in the gap, the edge states in the continuum, and the extended bulk states.}
    \label{fig5}
\end{figure}

The band structures (with topological index $\mathbb{Z}$) as a function of $\theta/\pi$ and wave function index, respectively, in Figs. \ref{fig5}(a) and \ref{fig5}(b) are shown with $L=5$ and $R=3$. There are $L(R-1)=10$ pairs of edge states in this case. The $L$ is odd, so one of the subsystems corresponding to $\lambda_l=0$ has chiral symmetry. In addition, the odd number of $R$ imposes that there is a band around zero energy. Also, there are $m-1=2$ pairs of edge states at zero energy. These result in the presence of zero-energy edge states in the continuum, see Fig. \ref{fig5}(a). As shown in Fig. \ref{fig5}(b), there are six pairs of edge states at finite energy in the continuum, represented by blue stars. Also, we have shown extended bulk states (two pairs edge states in the gap) by white (red) stars.

\section {Symmetry breaking perturbations} \label{s6}

 \begin{figure}[t!]
    \centering
    \includegraphics[width=1\linewidth]{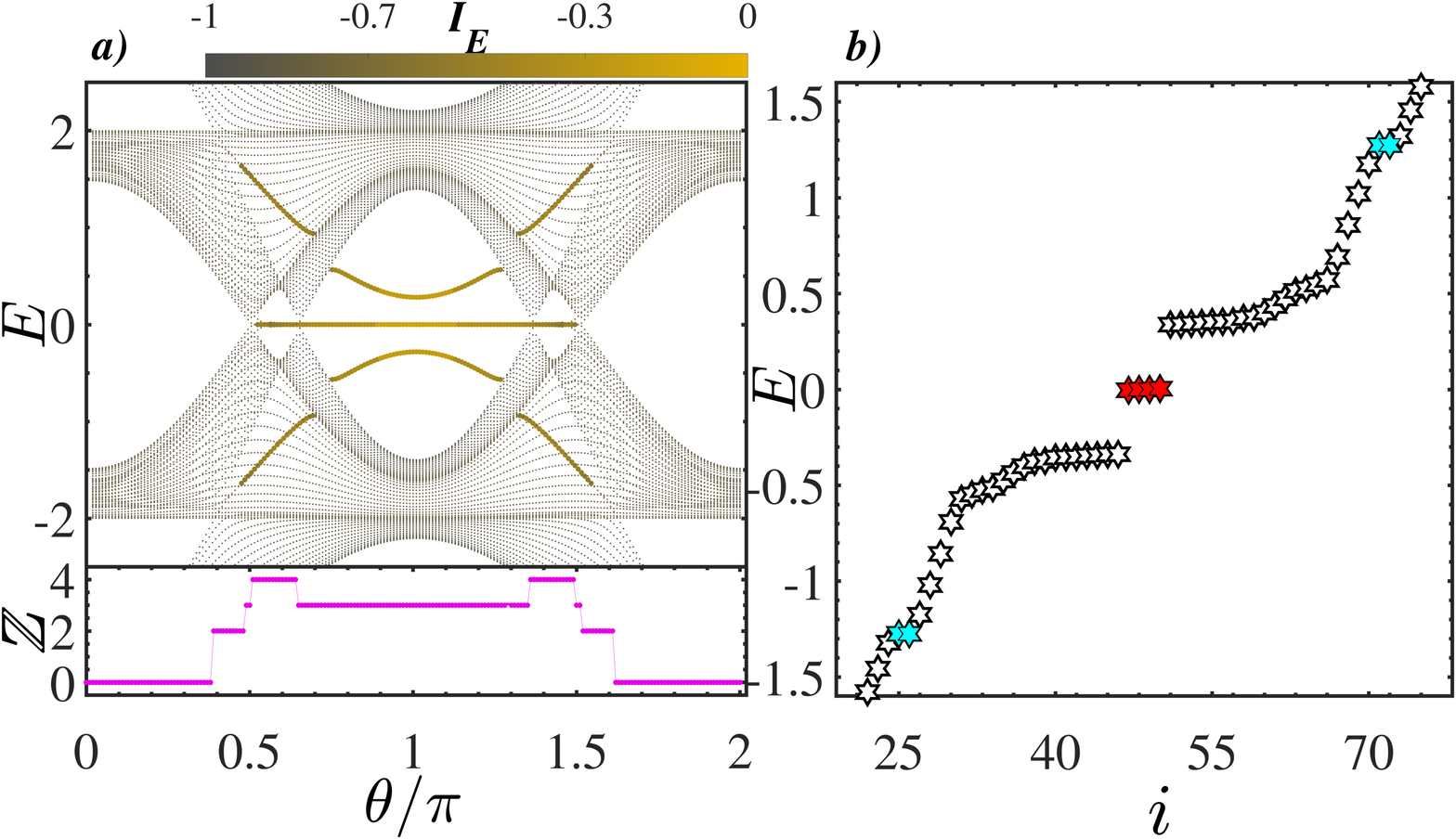}
    \caption{(Color online) (a) Dependence of
    energy spectra and relevant topological invariant on $\theta$ for $(L,R)=(3,2)$. Here the parameters are the same as before except that $t_{l=2}=0.1t$. (b) Energy spectrum as a function of wave function index corresponding to panel (a) with $\theta/\pi=0.6$. Red, blue and white markers represent, respectively, the edge states in the gap, the edge states in the continuum and the extended bulk states.}
    \label{fig6}
\end{figure}

To demonstrate the effect of subsidiary exchange symmetry on topological features of the system, we invoke a perturbation to break this symmetry. Without loss of generality, we investigate a special case, for an odd number of chains ($L=3, R=2$). The subsidiary exchange symmetry, having $\Upsilon_2$ operator, can be broken by setting either the intra- or inter-cell hopping amplitude of central chain different from the others, i.e., $t_{l=2}\ne t$ or $t_{l=2}^\prime \ne t^\prime$. Nevertheless, the system still has the main exchange symmetry with the $\Upsilon_1$ operator. After diagonalization, the system can be considered as two decoupled blocks, i.e., $SSH_R$ and $SSH_{2R}$. So there is no coupling term between the two subsystems and, interestingly, as shown in Fig. \ref{fig6}(a), the edge states of one block can lie within the bulk states of the other ones; the finite-energy edge states within $0.4\lesssim\theta/\pi\lesssim0.7$ and $1.3\lesssim\theta/\pi\lesssim1.6$. The breaking of the subsidiary exchange symmetry provides an off-diagonal term in one of the subsystems so that the edge states of that block can be hybridized with its bulk states resulting in the finite-energy delocalized topological edge states in the continuum; the finite-energy edge states within $0.7\lesssim\theta/\pi\lesssim0.75$ and $1.25\lesssim\theta/\pi\lesssim1.3$. Although, one may expect that the system has $L(R-1)=3$ edge states (including one pair at zero energy), another gap around zero energy is opened and an extra one pair of edge states manifests itself in this gap. Thus, in total, the system hosts two pairs of edge states at zero energy and at finite energy, giving rise to four pairs of edge states within $0.5\lesssim\theta/\pi\lesssim0.6$ and $1.4\lesssim\theta/\pi\lesssim1.5$, see also Fig. \ref{fig6}(b).
\begin{figure}[t!]
    \centering
    \includegraphics[width=1\linewidth]{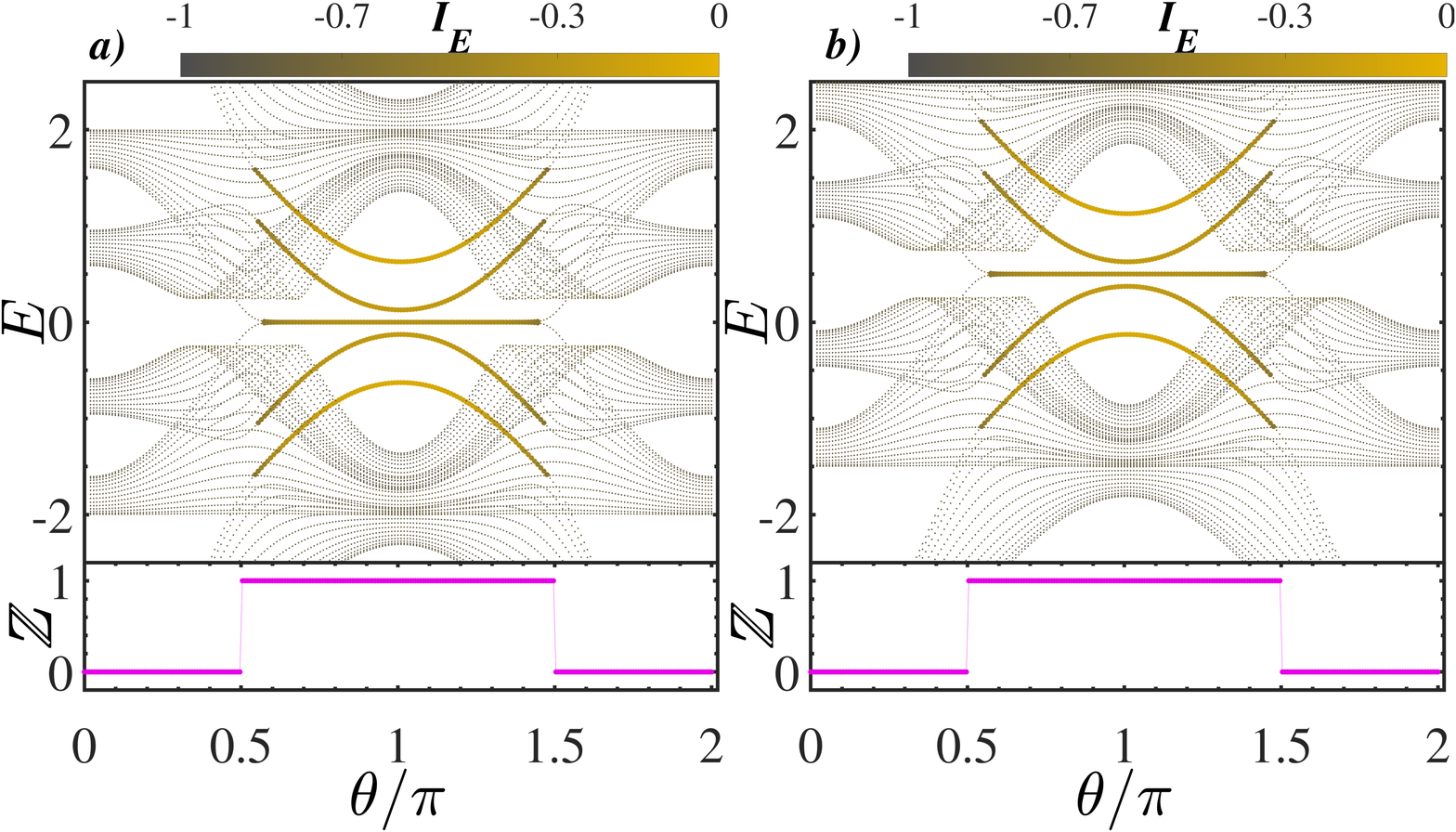}
    \caption{(Color online) Energy spectra and relevant topological invariant as a function of $\theta$ for $(L,R)=(3,2)$. (a) $V_1=0.5, V_2=0$ (b) $V_1=V_2=0.5$. By breaking the inversion symmetry of some subsystems and the whole system, the degeneracy of finite-energy edge states are destroyed and the gap around zero energy is opened. }
    \label{fig7}
\end{figure}

Finally, let us ensure that whether the inversion symmetry of subsystem is the fundamental symmetry protecting the edge states. To be more specific, we add the following perturbations
\begin{eqnarray}\label{e150}
H^{\prime}=\sum_{n=1}^{N}\sum_{l=1}^{3}\sum_{r=1}^{2} [(-1)^r V_1\delta_{l,2}+V_2] C_{n,l,r}^\dagger C_{n,l,r} ,
\end{eqnarray}
to Hamiltonian (\ref{RealHomil}) with parameters $(L,R)=(3,2)$. Here, the $\delta_{l,l^\prime}$ is the Kronecker delta and $V_1$ ($V_2$) is the amplitude of staggered (uniform) on-site potential breaking the inversion (chiral) symmetry of the full Hamiltonian. Despite breaking the subsidiary exchange symmetry, by rewriting the Hamiltonian in the basis of $\Upsilon_2$, one subsystem out of the three subsystems has inversion symmetry. While the other two subsystems lack the inversion symmetry. As shown in Fig. \ref{fig7}(a), for $V_1\neq0$ and $V_2=0$, one can see that one of the edge states is preserved due to presence of the inversion symmetry for the one of the subsystems. Whereas, the topology and the degeneracy of finite-energy of edge states, related to the two subsystems with broken inversion symmetry, are destroyed. Furthermore, turning on the $V_2$, breaking the full chiral symmetry, does not any effect on the subsystem-symmetry-protected edge states, as shown in Fig. \ref{fig7}(b). Consequently, importantly, the topology of the system is protected by the inversion symmetry of subsystems.

\section {Summary} \label{s7}

We considered the quasi-1D lattice system with $L$ legs and $R$ sublattices per leg. We determined symmetries of the system. It is shown that due to reflection and subsidiary inversion symmetry, the subsidiary exchange symmetry can be defined. Using this symmetry, we decomposed the system into subsystems. It is found the different topological phases can be emerged depending on $L$ and $R$. In the topological regime, each subsystem reveals $R-1$ topological edge states at zero and finite energy in the band structure. Consequently, there are, in total, $L(R-1)$ topological edge states. For the case with nonzero eigenvalues of the subsidiary exchange symmetry, the corresponding subsystems lack chiral symmetry and these subsystems belong to $AI$ class. But, when the axis of main exchange symmetry coincides on the central chain, due to vanishing one of the eigenvalues of subsidiary exchange symmetry, one of the subsystems would reduce to the original $SSH_R$ with chiral symmetry and this subsystem belongs to $BDI$ class. The existence of topological edge states does not depend on the symmetry of the whole Hamiltonian. Instead, the inversion symmetry of subsystems plays the key role in protecting the topological phases. Experimentally, the edges states can be observed in photonic lattices made of waveguide arrays \cite{Exper}.

{\it Acknowledgement}---.The authors would like to thank L.E.F. Foa Torres for helpful comments.

\begin{appendices}
\section{Allowed values of $l$ and their number} \label{AppxA}
In this Appendix, we determine which subsystems $l$ host the topological edge states at zero energy. In other word, the possible values for $l$ that can satisfy in Eq. (\ref{e15}) of the main text. Subsequently, we obtain the number of permissible value of $l$ for given $R$ and $L$.

Since Eqs. (\ref{e13}) and  (\ref{e130}) of the main text contain positive integers $R$, $2r$, $2r-1$, and $L$, so the expression $\frac{R l}{L+1}$ should be an integer. The maximum value of $l$ is $L$ so it is not possible to simplify this fraction, unless there exists a greatest common divisor (GCD) between the two values $(L+1,R)$. By introducing the GCD$(L+1,R)$ as $m$, the fraction can be rewritten as $\frac{R}{m}\frac{ml}{L+1}$. Because $\frac{R}{m}$ is a positive integer, $l$ should take the integer values $p\frac{L+1}{m}$ (with $p=1, 2, 3, ...$) to enforce the fraction $\frac{R}{m}\frac{ml}{L+1}$ to be a positive integer. 

Also $1 \le \ l \le L$ so 
\begin{eqnarray}
\frac{m}{L+1} \le \ p \le \ \frac{mL}{L+1},
\end{eqnarray}
where the lower bound is not integer and should be re-adjusted to $1$ according to the lower value of $p$. The integer upper bound of $p$ can be obtained as
\begin{eqnarray}
\frac{mL}{L+1}=m-\frac{m}{L+1}<m, 
\end{eqnarray}
subsequently,
\begin{eqnarray}
m-1 \le \ \frac{mL}{L+1}=m-\frac{m}{L+1}, 
\end{eqnarray}
resulting in
\begin{eqnarray}
1 \le \ p \le \ m-1.
\end{eqnarray}
So, the allowed values of $l$ are
\begin{eqnarray}
\frac{L+1}{m} \le \ l \le \ (m-1) \frac{L+1}{m},
\end{eqnarray}
and their number is $m-1$.
\end{appendices}


\end{document}